\documentclass[useAMS,usenatbib]{mn2e}
\usepackage{graphicx,amsmath}

\title[Pulse-wise Amati correlation]
  {Pulse-wise Amati correlation in Fermi GRBs}
\author[Basak \& Rao]{ Rupal Basak$^{1}$\thanks{E-mail: rupalb@tifr.res.in} and A. R. Rao$^{1}$\thanks{E-mail:
arrao@tifr.res.in}\\
$^{1}$Department of Astronomy and Astrophysics, Tata Institute
of Fundamental Research, Mumbai 400005, India}

\def\LaTeX{L\kern-.36em\raise.3ex\hbox{a}\kern-.15em
    T\kern-.1667em\lower.7ex\hbox{E}\kern-.125emX}

\begin{document}

\label{firstpage}

\maketitle

\begin{abstract}
We make a detailed pulse-wise study of gamma-ray bursts (GRBs) with known redshift
detected by \emph{Fermi}/Gamma Ray Burst Monitor (GBM). The sample contains 19 GRBs with 43 pulses.
We find that the average peak energy is correlated to the radiated energy (the Amati relation) for individual pulses 
with a correlation coefficient of 0.86, which is slightly better than the correlation for the full GRBs. 
As the present correlation holds within GRBs, it is a strong evidence supporting the reliability of
such a correlation. We investigate several aspects of this correlation. (i) We divide our sample into redshift 
bins and study the evolution of the correlation. 
Though there is a marginal indication of evolution of the correlation, we can conclude that the present data is 
consistent with no evolution. (ii) We compare the correlation in the first or single pulses of these GRBs 
to that of the rest of the pulses, and confirm that the correlation is unaffected by the fact that first/single pulses 
are generally harder than the rest. Finally, we conclude that the pulse-wise Amati correlation is 
more robust and it has the potential of refining the correlation so that GRB study could be used as a cosmological tool.

\end{abstract}


\section{INTRODUCTION}
Launched respectively in 2004 and 2008, \emph{Swift} and \emph{Fermi} jointly have become the primary workhorses for extensive studies of 
Gamma-ray bursts (GRBs). The Burst Alert Telescope (BAT) onboard \emph{Swift}, with its 5200 cm$^2$ detector area ($\sim1000$ cm$^2$ 
effective area) can detect very faint bursts down to $\rm 10^{-8}~ erg~cm^{-2}~s^{-1}$ (Barthelmy et al. 2005; Gehrels et al. 2004) in the 15-150 keV band. 
Due to rapid slew rate and high resolution instruments, \emph{Swift} has enabled redshift measurement of many GRBs. Fermi hosts two 
instruments, namely Gamma Ray Burst Monitor (GBM) and Large Area Telescope (LAT). GBM, the primary instrument for studying GRB 
prompt emission, covers a wide energy band (8 keV - 40 MeV; Meegan et al. 2009). Together \emph{Swift} and \emph{Fermi} have provided 
a wealth of prompt emission data. Swift has revealed that GRBs are frequent at very high redshifts (z). The highest spectroscopic 
z is 8.2 (Tanvir et al. 2009; Salvaterra et al. 2009), which corresponds to a very young universe, even less than 5\% of its present age. 
The high z of these objects make them valuable as high-z luminosity indicators. Indeed, correlations of some observables
(e.g., source frame peak energy --- $E_{\rm peak}$, lag --- $\tau$, variability etc.) with the energetics of a GRB (e.g., isotropic 
equivalent energy --- $E_{\rm \gamma, iso}$, isotropic peak luminosity --- $L_{\rm p, iso}$, collimation corrected energy 
--- $E_{\rm \gamma}$) have been found and discussed in the recent years (e.g., Fenimore \& Ramirez-Ruiz 2000; Norris et al. 2000; 
Amati et al. 2002; Schaefer 2003, 2004; Ghirlanda et al. 2004). In principle, through these correlations an observable can be
used to estimate the energetics and hence, the distance of the object. Therefore, it is suggested that GRBs can be used 
as ``standard candles'' in the same way as Type Ia supernovae (SNe). Moreover, the redshift barrier can be pushed to a much higher value
compared to SNe Ia.

These correlations are, however, empirical in nature, and may arise due to instrumental selection biases (e.g., see
Nakar \& Piran 2005; Band \& Preece 2005; Schaefer \& Collazzi 2007; Collazzi et al. 2012). One way to confirm  the reality of these 
correlations and understand  the effect of  selection biases is to examine the correlations within a GRB. For example, Ghirlanda et al. (2010),
using a set of 9 GRBs with known redshifts, have shown that the $E_{\rm peak}$ - $L_{\rm p, iso}$ correlation, known as the Yonetoku 
correlation (Yonetoku et al. 2004), holds within the time-resolved data. However, Basak \& Rao (2012a; BR12a hereafter), using the same set of GRBs,
have shown that the $E_{\rm peak}$ - $E_{\rm \gamma, iso}$ correlation (Amati correlation; Amati et al. 2002) breaks down in the 
time-resolved data. They concluded that the Amati correlation is not meaningful in time-resolved study. However, the fact that a GRB 
is made up of broad pulses (see e.g., Norris et al. 2005) and that the spectrum evolves independently in the individual pulses demand that 
pulses should be analyzed separately. BR12a used 22 pulses of these GRBs and found that the Amati correlation not only holds but 
it is better than the average Amati correlation.

In this paper, we expand the pulse-wise study of GRBs to include
  43 pulses of 19 GRBs. Our main motivation  is to establish that the pulse-wise correlation is suitable to 
use GRBs as standard candles for cosmological studies. Knowledge about the redshift evolution and other biases 
is very useful in this context. With a  larger sample compared to that used in BR12a, we can perform some critical tests. We first examine 
whether there is a redshift evolution of the pulse-wise Amati correlation by 
dividing our sample in various redshift bins. This is very crucial because the correlation has 
to hold for various redshifts if we want to use GRBs as ``standard candle'' at various redshifts.  
Another interesting feature to be investigated is the dependence of the pulse 
properties in their sequence of occurrence which might put additional constraints for using
a standard luminosity for a given pulse.  It has been suggested that GRBs tend to be harder at the beginning
than the rest of the burst (Crider et al. 1997; Ghirlanda et al. 2003; Kaneko et al. 2003; Ryde and Pe'er 2009). 
This hardness should reflect in the peak energy of the spectrum. For a given energy, the peak energy of the first pulses should be 
biased towards higher values and will tend to give a systematic shift in the peak energy luminosity correlation.
 We examine this possibility by dividing our sample into first/single pulses and rest of the pulses, 
and then studying the pulse-wise Amati correlation for them. 

Following is the plan of this paper. In Section 2, we shall present our 
sample and the analysis technique. Results are discussed in Section 3. Major conclusions are discussed in Section 4.

\begin{table*}

\caption{The observer frame peak energy ($E_{\rm peak,obs}$) and the isotropic energy ($E_{\gamma,iso}$)
for individual pulses of the 10 GRBs. For the rest of 9 GRBs, see Basak \& Rao (2012a) }

\begin{tabular}{cccccccccc}
\hline 

GRB & z & Pulse & $t_{\rm 1}$(s) & $t_{\rm 2}$(s) & $\alpha$ & $\beta$ & $E_{\rm peak,obs}$ (keV)& $\chi_{red}^{2}$ (dof) & $E_{\rm \gamma,iso}$ ($10^{52}$erg) \\
\hline
\hline 
090902B & 1.822 & 1 & 5.0 & 13.0 & $-0.23_{-0.13}^{+0.13}$ & $-3.56_{-0.56}^{+0.22}$ & $828.9_{-28.7}^{+31.6}$ & 1.03 (243) & 178.24 \\
        &       & 2 & 12.0 & 18.0 & $-0.76_{-0.04}^{+0.07}$ & $-3.21_{-0.38}^{+0.23}$ & $537.3_{-23.7}^{+23.5}$ & 1.34 (290) & 111.81 \\
        &       & 3 & 18.0 & 23.0 & $-0.76_{-0.03}^{+0.04}$ & $-2.44_{-0.11}^{+0.08}$ & $285.4_{-17.0}^{+15.6}$ & 1.34 (275) & 62.14 \\
\hline
090926A & 2.1062 & 1 & 0.0 & 8.0 & $-0.55_{-0.02}^{+0.02}$ & $-2.44_{-0.05}^{+0.05}$ & $332.1_{-9.6}^{+9.7}$ & 1.59 (492) & 116.83 \\
        &       & 2 & 8.0 & 15.0 & $-0.80_{-0.02}^{+0.02}$ & $-2.90_{-0.18}^{+0.13}$ & $241.3_{-7.3}^{+7.6}$ & 1.55 (466) & 61.15 \\
\hline
090926B & 1.24 & 1 & 12.0 & 65.0 & $0.13_{-0.40}^{+0.45}$ & $-10.0$ & $73.8_{-6.1}^{+7.7}$ & 0.65 (42) & 2.54 \\
\hline
091003A & 0.8969 & 1 & 13.95 & 26.24 & $-0.953_{-0.06}^{+0.07}$ & $-2.38_{-0.51}^{+0.20}$ & $299.4_{-41.0}^{+48.2}$ & 1.10 (293) & 6.05 \\
\hline
091020  & 1.71 & 1 & -2.0 & 15.0 & $-1.16_{-0.15}^{+0.22}$ & $-2.07_{-0.50}^{+0.24}$ & $197.3_{-75.6}^{+115.9}$ & 1.07 (134) & 8.0 \\
\hline
091024  & 1.092 & 1 & -7.94 & 33.02 & $-0.95_{-0.14}^{+0.22}$ & $-2.08_{-\infty}^{+0.47}$ & $725.0_{-162.8}^{+226.7}$ & 0.82 (119) & 9.10 \\
        &       & 2 & 200.71 & 249.86 & $-0.81_{-0.26}^{+0.40}$ & $-9.37$ & $112.7_{-14.6}^{+14.4}$ & 1.43 (81) & 3.74 \\
        &       & 3 & 313.35 & 346.12 & $-1.18_{-0.07}^{+0.10}$ & $-9.36$ & $225.7_{-27.5}^{+19.4}$ & 1.60 (335) & 9.14 \\
        &       & 4 & 622.7 & 664.7 & $-1.17_{-0.07}^{+0.07}$ & $-2.15$ & $371.0_{-71.0}^{+111.0}$ & 1.09 (473) & 2.45 \\
\hline
091127  & 0.490 & 1 & -2.0 & 4.0 & $-0.92_{-0.16}^{+0.19}$ & $-2.20_{-0.14}^{+0.08}$ & $65.8_{-9.3}^{+12.1}$ & 1.35 (151) & 1.06 \\
        &       & 2 & 5.0 & 14.0 & $-1.34_{-0.33}^{+0.77}$ & $-2.88_{-0.17}^{+0.17}$ & $14.6_{-3.5}^{+1.7}$  & 1.00 (140) & 0.44 \\
\hline
091208B  & 1.063 & 1 & -1.0 & 5.0  & $-1.36_{-0.24}^{+1.08}$ & $-2.30$ & $74.2_{-34.9}^{+41.2}$ & 1.19 (154) & 0.55 \\
         &       & 2 &  6.0 & 13.0 & $-1.25_{-0.13}^{+0.13}$ & $-2.84_{-\infty}^{+0.48}$ & $113.7_{-15.8}^{+30.8}$  & 1.19 (223) & 1.31 \\
\hline
100414A & 1.368 & 1 & 1.0  & 13.0 & $-0.14_{-0.07}^{+0.08}$ & $-4.90_{-\infty}^{+1.47}$ & $557.5_{-28.5}^{+31.1}$ & 1.10 (275) & 22.73 \\
        &       & 2 & 14.0 & 20.0 & $-0.56_{-0.06}^{+0.06}$ & $-3.52_{-\infty}^{+0.71}$ & $599.4_{-44.3}^{+49.7}$ & 1.01 (238) & 14.26 \\
        &       & 3 & 21.0 & 28.0 & $-0.91_{-0.05}^{+0.06}$ & $-2.76_{-2.42}^{+0.39}$ & $635.1_{-78.3}^{+93.5}$ & 1.17 (240) & 12.00 \\
\hline
100814A  & 1.44  & 1 & -3.0 & 5.0  & $1.04_{-0.50}^{+0.65}$ & $-3.00_{-2.55}^{+0.71}$ & $168.6_{-22.1}^{+25.8}$ & 0.92 (176) & 2.14 \\
         &       & 2 &  4.0 & 14.0 & $0.84_{-0.36}^{+0.55}$ & $-3.43_{-\infty}^{+0.95}$ & $133.5_{-16.42}^{+13.8}$  & 0.79 (130) & 2.32 \\
\hline

\end{tabular}
\label{table1}

\end{table*}

\section{Data selection and Analysis}
The set of GRBs analyzed in the previous work (see BR12a) was taken from Ghirlanda et al. (2010) which reports 9 GRBs from 
2008 to 2009 detected by \emph{Fermi}/GBM, the last one being GRB 090618 (i.e., 2009 June). In this paper,
we have analyzed all GRBs with known redshift from June 2009 till August 2010. We have then combined these two sets
for correlation analysis. Redshift 
measurement is essential to measure $E_{\rm \gamma, iso}$. The set is taken from the web-page of Jochen 
Greiner ($\rm http://www.mpe.mpg.de/\sim jcg/grbgen.html$). Three of these GRBs have been detected first by Fermi, and the rest 
of them were detected first by Swift.

In these GRBs, we carefully choose broad pulses, as described in BR12a. The pulses must not have large overlap with a rapidly
varying profile. We choose only smooth pulses, or a portion of a pulse (see BR12a for details of pulse selection criteria). 
In Table~\ref{table1}, we show the start and stop time of a given pulse. Spectra are generated by integrating over these time bins, 
and then re-binned in energy, with NaI re-binned with the requirement of $\sim40$ counts per bin and BGO detectors with $\sim$50 
counts per bin. We generally take two NaI and one BGO detector for spectral analysis. If a third NaI detector has
comparable counts, then we also include that detector for spectral analysis. An effective area correction is applied for 
these detectors. We fit the spectrum in XSPEC version 12.6.0 by $\chi^2$ minimization. For uniformity, we model all the 
spectra by Band model (Band et al. 1993), except for GRB 090902B, for which we use a power-law along with the Band model.
We determine the parameters of the model with nominal 90\% error. The Band model is a phenomenological representation of
GRB emission, and is a smoothly joined broken power-law with four parameters, namely, 
$\alpha$, $\beta$ as power-law indices, $E_{\rm peak,obs}$ as the observer frame peak energy (in EF(E) representation) 
and $N_{\rm b}$ as the normalization. The source frame peak energy ($E_{\rm peak}$) is calculated by multiplying the 
observed peak energy by (1+z). $E_{\rm \gamma, iso}$ is calculated for each pulse in the 1 to 10000 keV band in the source 
frame. The NaI and BGO detectors cover portions of this energy range. Hence, we calculate the flux by extrapolating 
the best-fit model in the said energy band.

In BR12a, the pulse description of Basak \& Rao (2012b) was used and it was found that the replacement of $E_{\rm peak}$
with another parameter of the model, namely the peak energy at zero fluence ($E_{\rm peak,0}$), makes the correlation stronger.
This model assumes that $E_{\rm peak}$ evolves with ``running fluence'' as an exponential law given by 
$E_{\rm peak}(t) =E_{\rm peak,0}~ \rm exp-(\frac{\phi(t)}{\phi_0})$, where $\phi(t)$ is the ``running fluence'' at time `t',
defined as the integrated flux from the start of the poulse till time `t'. (Liang \& Kargatis 1996).
However, it is known that the $E_{\rm peak}$ evolution can be more complicated than the simple assumption of Basak \& Rao (2012b)---
there may be ``intensity tracking'' nature in the evolution (e.g., Ford et al. 1995; Liang \& Kargatis 1996; Kaneko et al. 2006; 
Lu et al. 2010; Hakkila \& Preece 2011; Lu et al. 2012; Basak \& Rao 2013a; b). Hence, in the present paper 
we restrict ourselves only to the pulse-wise Amati correlation.

To study the $E_{\rm peak}$ - $E_{\rm \gamma, iso}$ pulse-wise correlation we use Pearson linear correlation (r) and Spearman rank 
correlation ($\rho$), with their corresponding chance probability ($P_{\rm r}$ and $P_{\rm \rho}$ respectively). The Pearson 
correlation is calculated by taking logarithmic values of the measured quantities. We fit the scatter plots by a linear model 
of the form: $\rm log(\frac{E_{\rm peak}}{100 \rm keV})=K+\delta \rm log(\frac{E_{\rm \gamma, iso}}{10^{52} \rm erg})$. We assume
an intrinsic scatter ($\sigma_{\rm int}$) for the $E_{\rm peak}$ following BR12a (see D'Agostini 2005). We maximize a 
joint likelihood function ($L$) following Eqn. 17 of Wang et al. (2011) and find the $\chi^2$ by $\chi^2$=-2ln$L$
(see Eqn. 24 of Wang et al. 2011).

For all analysis, we assume $\Lambda$-CDM cosmology and use the following cosmological parameters. Hubble parameter, 
$H_{\rm 0}=\rm 71~km ~s^{-1}~Mpc^{-1}$, dark energy density, $\Omega_{\rm \Lambda}=0.73 $, total baryonic and dark 
matter density, $\Omega_{\rm m}=0.27$, which implies a spatially flat universe. These values are determined combining 
the data of seven-year Wilkinson Microwave Anisotropy Probe (WMAP) observations (Jarosik et al. 2011), Baryon Acoustic 
Oscillations (Percival et al. 2010) and Type Ia supernova (Riess et al. 2011). 
Note that the recent measurements by Planck mission reports slightly different
values of these parameters: $H_{\rm 0}=\rm 67.3~km ~s^{-1}~Mpc^{-1}$, $\Omega_{\rm m}=0.315$ (Planck Collaboration, et al. 2013).
These new values will have 
a marginal effect on $E_{\rm \gamma, iso}$, about 7\% at low redshifts
(z = 1) and about 4\% at high redshift (z = 3) 
and hence, the slope ($\delta$) and normalization (K) of the pulse-wise Amati correlation will not be affected significantly.

\begin{table*}\centering
\caption{Results of linear fit to the logarithmic values of $E_{peak}$-$E_{\gamma,iso}$ data for the GRBs. K, $\delta$ and 
$\sigma_{int}$ are the linear fit parameters: K is intercept, $\delta$ is slope and $\sigma_{int}$ is the intrinsic scatter 
of the data. The $E_{peak}$ and $E_{\gamma,iso}$ are normalized to 100 keV and $10^{52}$ erg, respectively.
}

 \begin{tabular}{c|c|c|c|c}
\hline
 Data set & K & $\delta$ & $\sigma_{int}$ & $\chi^2_{red}$ (dof)\\
\hline
\hline 
All & $0.269\pm0.041$ & $0.499\pm0.035$ & $0.256\pm0.0344$ & 1.04 (41) \\
z\textless1 bin & $0.223\pm0.077$ & $0.523\pm0.113$ & $0.306\pm0.065$ & 1.13 (15) \\
1\textless z\textless2 bin & $0.391\pm0.056$ & $0.439\pm0.048$ & $0.208\pm0.047$ & 1.10 (14) \\
z\textgreater2 bin & $0.373\pm0.068$ & $0.421\pm0.038$ & $0.200\pm0.063$ & 1.24 (8) \\
\hline
\hline
First/Single pulses & $0.352\pm0.055$ & $0.465\pm0.044$ & $0.221\pm0.048$ & 1.10 (17) \\
Rest of the pulses & $0.228\pm0.057$ & $0.503\pm0.050$ & $0.273\pm0.048$ & 1.09 (22) \\

\hline
\end{tabular}
\label{table2}
\end{table*}

\begin{figure}
\centering
\includegraphics[width=3.0in]{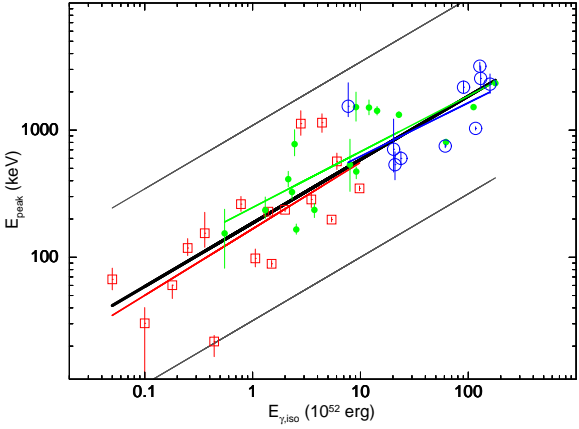} 
\caption{Amati correlation for the 43 pulses of 19 GRBs. The thick line shows the linear fit to the log($E_{\rm peak}$)-
log($E_{\rm \gamma, iso}$) data. The parallel thin lines are 3$\sigma_{\rm int}$ scatter of the data. The other thin lines 
are linear fit for the data in various redshift bins. Markers are: boxes for z\textless1, filled circles 
for 1\textless z\textless2 and open circles for z\textgreater2 bins.
\label{fig1}}
\end{figure}

\begin{figure}\centering
{

\includegraphics[width=3.0in]{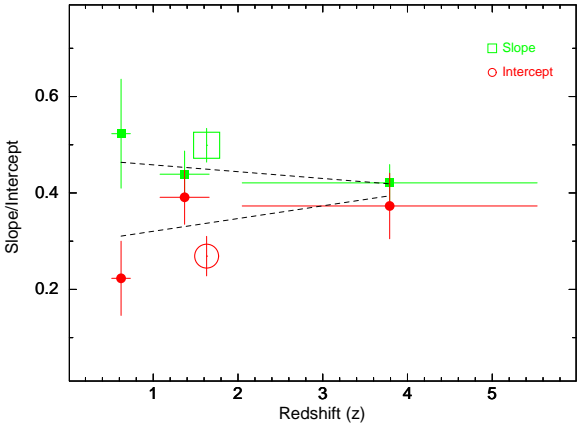} 

}
\caption{Redshift evolution of the pulse-wise Amati correlation. The values of the slope ($\delta$) and intercept (K)
in different redshift bins are shown by filled boxes and filled circles. The average values of $\delta$ and K are 
shown by similar open symbols. The data are fitted by straight lines to calculate approximate evolution of the correlation.
}
\label{fig2}
\end{figure}

\begin{figure}\centering
{

\includegraphics[width=3.0in]{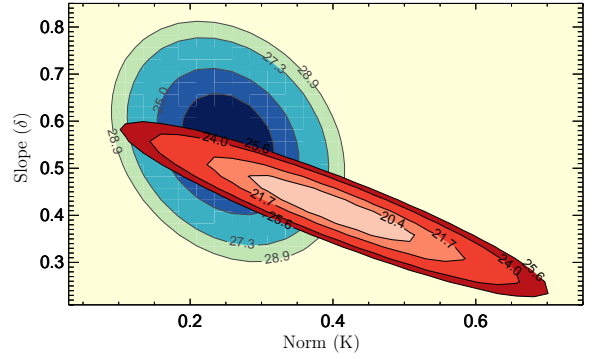} 

}
\caption{$\chi^2$ contour plot of K and $\delta$ for two different redshift bins --- z$\leqslant 1.092$ 
(blue contours in the left) and z\textgreater1.092 (red contours in the right).
Contour levels in both cases are $\Delta \chi^2=1.0, 2.3,4.61,6.17$ corresponding to 1$\sigma$ for 1 parameter, 
1$\sigma$, 2$\sigma$ and 3$\sigma$ for 2 parameters. Note that the values are matching within $\Delta \chi^2=1.0$.
}
\label{c1}
\end{figure}

\section{Results}

\subsection{Pulse-wise Amati correlation}
In Table~\ref{table1} we show our set of GRBs in chronological order. The redshifts are quoted from the web page of
Jochen Greiner. In columns 4 and 5 we specify the time interval for a given pulse. In the next three columns we show the parameters 
of Band model fit. Note that the low energy photon index ($\alpha$) sometime crosses the ``synchrotron line of death'' of electron slow 
cooling regime (i.e., the value of $\alpha$ is greater than -2/3; Preece et al. 1998). Hence, these models are either 
unphysical or the spectrum is not synchrotron (Zhang \& Meszaros, 2002). However, the $\chi^2_{\rm red}$ of the fits 
are acceptable. We are interested 
in measuring the average properties of the pulses, namely, peak of the EF(E) spectrum and the bolometric flux. Looking at 
the $\chi^2_{\rm red}$ of the fits we can safely say that these values should not vary much for any other model which gives 
acceptable fit. The isotropic energy for the individual pulses are shown in the last column.

In Figure~\ref{fig1}, we show the scatter plot of $E_{\rm peak}$ vs. $E_{\rm \gamma, iso}$. The correlation coefficients are
as follows: r=0.86 with $P_{\rm r}=1.50\times10^{-13}$, $\rho$=0.86 with $P_{\rm \rho}=7.47\times10^{-14}$. If we do not use the
logarithmic values then the Pearson correlation is 0.80. We note that there is a significant correlation in pulse-wise Amati data. 
Note that due to larger sample with respect to BR12a 
the significance of the correlation has improved (previously $P_{\rm \rho}$ was $4.57\times 10^{-8}$). To obtain the average Amati
correlation we use the data for these GRBs, published by Gruber et al. (2011). The correlation coefficients are r=0.83 with 
$P_{\rm r}=1.10\times10^{-5}$, $\rho$=0.85 with $P_{\rm \rho}=4.27\times10^{-6}$, and r without taking logarithmic values 
is 0.69 with $P_{\rm r}=1.08\times10^{-3}$. 

The thick line in Figure~\ref{fig1} is the linear fit to the logarithmic values of pulse-wise Amati data. The corresponding 
fit results are shown in Table~\ref{table2} (first row). Note that the slope 
($\delta$) of the linear fit is $\sim0.5$, which gives an empirical relation as $E_{\rm peak} \sim E_{\rm \gamma, iso}^{1/2}$.
The value of the slope ($\delta$) and the intercept (K) for 22 pulses, reported by BR12a, were $0.516\pm0.049$ and $0.289\pm0.055$.
The current values are: $0.499\pm0.035$ and $0.269\pm0.041$. $\delta$ is within 0.36$\sigma$ and K is within 0.36$\sigma$ of the
previous values. Note that due to larger size of the present data, the values are now determined with better accuracy. However, the 
intrinsic scatter ($\sigma_{\rm int}$) has increased to $0.256\pm0.0344$ from $0.244\pm0.048$, but still within 0.25$\sigma$
of the old value. In Figure~\ref{fig1}, the lines parallel to the thick line show the 3$\sigma_{\rm int}$ scatter of the data.

\subsection{Redshift evolution of the correlation}
In order to investigate the possible evolution with redshift of the pulse-wise $E_{\rm peak}$-$E_{\rm \gamma, iso}$ correlation,
we divide our data set in three redshift bins: (i) z\textless 1 (contains 17 data points), (ii) 1\textless z \textless2 (16 data points) 
and (iii) z\textgreater3 (10 data points). 
In Figure~\ref{fig1}, we show these different data by different markers. We fit the individual scattered data by a straight line
as before. The thin lines represent these fits. In Table~\ref{table2} (2nd, 3rd and 4th row), we show the K,$\delta$, 
$\sigma_{\rm int}$ and $\chi^2_{\rm red}$ of these fits. We note that there are deviations of these lines from the average correlation line.
However, the sample size is too low at present to conclude that these variations are significant. In figure~\ref{fig2},
we show the slope ($\delta$; shown by filled boxes) and the intercept (K; shown by filled circles) of these linear fits 
as functions of z. As there are only three redshift bins, we can fit at most a straight line to the z-$\delta$ and z-K data. 
The dashed lines in Figure~\ref{fig2} show the linear fits to the data. The slope of the z-$\delta$ fit is 
$\frac{d \delta}{dz}=(-1.41\pm3.74)\times10^{-2}$, while that of the z-K fit is $\frac{dK}{dz}=(2.64\pm4.88)\times10^{-2}$.
The errors quoted are nominal 90\% errors. The average values of K and $\delta$ are shown by open circle and open box respectively.

We note that there is an indication of evolution of the correlation. However, the errors in $\frac{dK}{dz}$ and $\frac{d \delta}{dz}$
are large, and hence, the values are not statistically significant. Redshift evolution of Amati correlation and other prompt emission 
correlations have been studied by some researchers (Li 2007; Tsutsui et al. 2008; Ghirlanda et al. 2008; Azzam 2012). They have 
reported mild evolution of the correlation. Note that the evolution of Tsutsui et al. (2008) is monotonic, but Figure~\ref{fig2} 
indicates that the there is no systematic evolution of the pulse-wise Amati correlation. 

To further investigate this matter, we divide our sample into low and high redshift bins. As there are different number of pulses in a given
GRB, the division is not exactly half. The lower redshift bin contains 23 pulses upto redshift 1.096, and the other sample contains the
rest. In Figure~\ref{c1}, we have shown the $\chi^2$ contour of K and $\delta$ for these two redshift bins. The contour levels are 1$\sigma$ 
for single parameter (i.e., $\Delta \chi^2 = 1.0$), 1$\sigma$, 2$\sigma$ and 3$\sigma$ for two parameters (i.e, $\Delta \chi^2 = 2.3$,
4.61 and 6.17, respectively). Note that they agree within $\Delta \chi^2 = 1.0$. Hence, we conclude that the present data is still consistent 
with no evolution, within statistical limit. Yonetoku et al. (2010),
for example, have studied both Amati and Yonetoku correlation in a set of 101 GRBs from different detectors. They found that the
Yonetoku correlation has a redshift evolution at 2$\sigma$ level, while the Amati correlation has that at 1$\sigma$ level.
Hence, the evolution of both Amati and pulse-wise Amati correlations are very weak and insignificant.

\begin{figure}\centering
{

\includegraphics[width=3.0in]{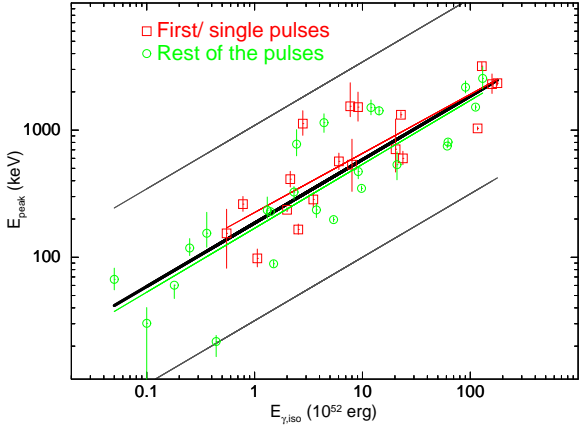} 

}
\caption{Amati correlation for 19 first/single pulses (open boxes) and 24 rest of the pulses (open circles).
The parallel straight lines are the same as in Figure~\ref{fig1}
}
\label{fig3}
\end{figure}

\begin{figure}\centering
{

\includegraphics[width=3.0in]{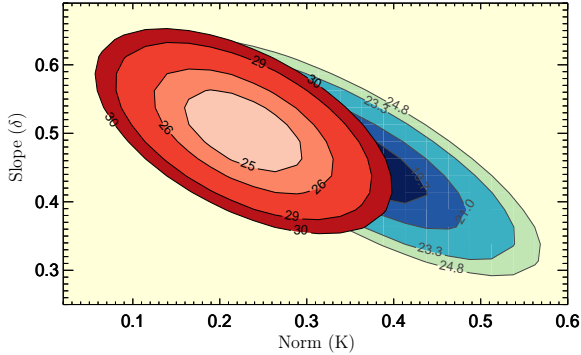} 

}
\caption{$\chi^2$ contour plot of K and $\delta$ for first/single (blue contours in the right) and rest of the pulses
(red contours in the left). The contour levels 
are the same as in Figure~\ref{c1}. Note that the values are matching within $\Delta \chi^2=1.0$.
}
\label{c2}
\end{figure}

\begin{figure}\centering
{

\includegraphics[height=3.2in, width=2.0in, angle=-90]{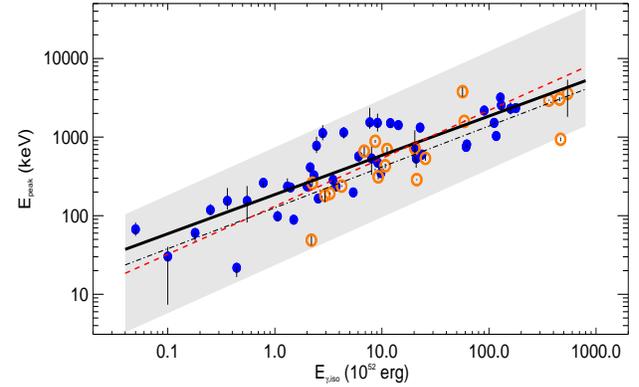} 

}
\caption{Data set of the present paper (filled circle) plotted on the time-integrated $E_{\rm peak}$-$E_{\rm \gamma, iso}$ plane of 
the complete sample of Nava et al. (2012). The dashed line shows the linear fit to the logarithmic time-integrated data, while the shaded 
region shows the 3$\sigma_{int}$ of the data (from Nava et al. 2012). The linear fit to the logarithmic pulse-wise data is shown by a thick 
line for comparison. In the same plot, time-integrated data of the set of 19 GRBs used for our analysis are also shown (open circles). The 
best fit line to the logarithmic values of the data of these GRBs is shown by dot-dashed line.
 }
\label{compare}
\end{figure}

\subsection{Trend of the single/first pulses}
We divide our sample into single/first pulses (19 data points) and the rest of the pulses (24 data points). If the first/single pulses
have harder spectrum, the peak energy for a given luminosity should have a bias towards higher value. We fit linear curves to the 
log($E_{\rm peak}$) - log ($E_{\rm \gamma, iso}$) data for both the cases. The corresponding fit parameters are reported in 
Table~\ref{table2}. Figure~\ref{fig3} shows the correlation for first/single (open boxes) and the rest of the pulses (open circles). 
The average linear fit to the full data is over-plotted as thick line. Note that the intercept (K) of the first/single pulses 
($0.352\pm0.055$) is higher than the rest of the pulses ($0.228\pm0.057$). However, the slope ($\delta$) is lower ($0.465\pm0.055$)
than the rest ($0.503\pm0.057$). Note that $E_{\rm peak}~\sim~10^K\times E^{\delta}_{\rm \gamma, iso}$, hence the bias is nullified 
by a combination. To investigate this further, we draw $\chi^2$ contour plot of K and $\delta$ for the two cases
(see Figure~\ref{c2}). The blue contours at the right are those for the single/first pulses. The contour levels are the same as in
Figure~\ref{c1}. Note that the two sets agree with each other within 1$\sigma$. Hence, we conclude
that the correlation is unaffected by the fact that the first/single pulses are harder.

\subsection{Comparison with time-integrated correlation}
In order to compare the pulse-wise Amati correlation with the familiar time-integrated Amati correlation, we use the recent
results of Nava et al. (2012). They have studied the time-integrated spectrum of 136 GRBs with known redshift.
We denote the slope as $\delta_1$ and the normalization by $K_1$ for their value to distinguish them from the present work.
From Nava et al. (2012) we quote the following values. (a) For the full set of 136 GRBs $\rho=0.77$, $\delta_1=0.55\pm0.02$,
$K_1=-26.74\pm1.13$, $\sigma_{int}=0.23$. (b) For a complete set of 46 GRBs $\rho=0.76$, $\delta_1=0.61\pm0.04$,
$K_1=-29.60\pm2.23$, $\sigma_{int}=0.25$, (c) for the complementary sample of 90 GRBs, $\rho=0.78$, $\delta_1=0.531\pm0.02$,
$K_1=-25.63\pm1.35$, $\sigma_{int}=0.25$.

In Figure~\ref{compare}, we have shown together the data points of the pulse-wise analysis and the time-integrated 
$E_{\rm peak}$-$E_{\rm \gamma, iso}$ correlation of the complete sample (Nava et al. 2012). We have shown the 3$\sigma_{int}$ scatter in the 
time-integrated data by a  shaded region. 
In our analysis, we have normalized the $E_{\rm peak}$ by 100 keV and $E_{\rm \gamma,iso}$ by $10^{52}$ erg. Converting this value to
be compatible with the Nava et al. (2012), we obtain $K=-23.68 \pm 1.82$, which is comparable those obtained by time-integrated 
analysis. The slope ($\delta$) of the time-integrated correlation ($0.55\pm0.02$) is also comparable to the pulse-wise correlation 
($0.499\pm0.035$). To compare these correlations, we have plotted the time-integrated correlation line (dashed line) and 
the pulse-wise correlation line (thick line). In order to find where the time-integrated data of the 19 GRBs used for 
the present study lie, we also show the time-integrated values (taken from David Gruber --- private communication) by open circles. 
The linear fit to this time-integrated data is shown by dot-dashed line.

Note that, all these lines are comparable. Though the pulse-wise Amati correlation is comparable to the time-integrated 
correlation in terms of correlation coefficient, compared to the time-integrated correlation of Nava et al. (2012) pulse-wise Amati 
correlation is tighter ($\rho=0.86$). This points to the fact that pulse-wise correlation is indeed meaningful.

\section{  Discussion and  Conclusions}

To summarize, we have enlarged our sample twofold compared to BR12a to study the pulse-wise Amati correlation. We confirm that the correlation holds
with a good correlation coefficient. We could divide our sample in various redshifts and study the same correlation in these redshift bins.
We found no statistically significant evolution of the correlation with redshift. Possibly a larger sample is needed to further
investigate this matter. To check whether the correlation has a bias due to the fact that the first/single pulse have a tendency 
for higher peak energy ($E_{\rm peak,obs}$), we divided our sample into the first/single pulses and the rest of the pulses. 
We confirm that the correlation remains unaffected by such a bias. 

Another interesting fact we notice from close inspection of Figure~\ref{compare} is as follows.
The time-integrated and pulse-wise correlation are in general agreement with each other (compare the thick line
with the dashed line). However, the best fit line (dot-dashed) to the time-integrated data of 19 GRBs (present sample) lies 
slightly lower than that of the pulse-wise data. Though the slope of these lines are similar, their normalization is slightly different.
The difference in the normalizations of the pulse-wise correlation and the time-integrated
correlation can be understood as due to the summing of the individual pulses
in the individual GRBs. For example, if a GRB has two pulses, the time averaged
value of E$_{\rm peak}$ would be the average value of individual  E$_{peak}$ values of 
the pulses (possibly biased towards the brighter pulse), whereas the value of 
$E_{\rm \gamma, iso}$ would simply be the sum of individual pulse values. Hence
we can expect a shift in the data point towards right by about a factor of two.
However, a larger sample is required to draw such conclusion, but at present,
there is no significant evidence of different slope, dispersion and normalization between
the ``two versions'' of the correlation.

The present work strongly indicates that the Amati correlation is a pulse characteristics
rather than an average GRB property. The strengthening of the Amati correlation when
taken pulse-wise is a clear vindication that the correlation is real and not due
to any artifact of selection or observational biases. It would be interesting
to investigate all GRB prompt emission correlations taking individual 
pulses so that a very tight multi-parameter correlation could be obtained 
which will eventually lead to the use of GRBs as standard candles for
cosmology.

\section*{Acknowledgments} This research has made use of data obtained through the
HEASARC Online Service, provided by the NASA/GSFC, in support of NASA High Energy
Astrophysics Programs. 
RB would like to thank David Gruber for providing useful information
of \emph{Fermi}/GBM GRBs. A sincere thank to David Fanning (Fanning Software Consulting, Inc.)
for online help of IDL plots. We thank the referee for valuable suggestions which
 improved the 
quality of the paper.

\end{document}